\documentclass[%
 twocolumn,
 reprint,
superscriptaddress,
 amsmath,amssymb,
 aps,
]{revtex4-1}

\usepackage{graphicx}
\usepackage{dcolumn}
\usepackage{bm}
\usepackage{xcolor}
\usepackage[normalem]{ulem}

\usepackage{hyperref}
\hypersetup{
 linktocpage = true,
     colorlinks,
    citecolor=blue,
    filecolor=black,
    linkcolor=blue,
    urlcolor=blue
}

\begin{document}


\title{The learnability scaling of quantum states: restricted Boltzmann machines}

\author{Dan Sehayek}
\affiliation{%
 Department of Physics and Astronomy, University of Waterloo, Ontario, N2L 3G1, Canada}
 \affiliation{
 Perimeter Institute for Theoretical Physics, Waterloo, Ontario N2L 2Y5, Canada
 }
 \author{Anna Golubeva}
\affiliation{%
 Department of Physics and Astronomy, University of Waterloo, Ontario, N2L 3G1, Canada}
 \affiliation{
 Perimeter Institute for Theoretical Physics, Waterloo, Ontario N2L 2Y5, Canada
 }
 \author{Michael S. Albergo}
\affiliation{%
Department of Physics and Astronomy, University of Waterloo, Ontario, N2L 3G1, Canada}
\affiliation{
Perimeter Institute for Theoretical Physics, Waterloo, Ontario N2L 2Y5, Canada
 }
  \author{Bohdan Kulchytskyy}
\affiliation{%
 Department of Physics and Astronomy, University of Waterloo, Ontario, N2L 3G1, Canada}
 \affiliation{
 Perimeter Institute for Theoretical Physics, Waterloo, Ontario N2L 2Y5, Canada
 }
\author{Giacomo Torlai}
\affiliation{%
Center for Computational Quantum Physics, Flatiron Institute, New York, New York, 10010, USA
}%
\author{Roger G. Melko}
\affiliation{%
 Department of Physics and Astronomy, University of Waterloo, Ontario, N2L 3G1, Canada}
 \affiliation{
 Perimeter Institute for Theoretical Physics, Waterloo, Ontario N2L 2Y5, Canada
}

\date{\today}

\begin{abstract}
Generative modeling with machine learning has provided a new perspective on the data-driven task of reconstructing quantum states 
from a set of qubit measurements.  As increasingly large experimental quantum devices are built in laboratories,
the question of how these machine learning techniques scale with the number of qubits is becoming crucial.
We empirically study the scaling of restricted Boltzmann machines (RBMs) applied to reconstruct ground-state wavefunctions of 
the one-dimensional transverse-field Ising model from projective measurement data.
We define a learning criterion via a threshold on the relative error in the energy estimator of the machine.  
With this criterion, we observe that the number of RBM weight parameters required for accurate 
representation of the ground state in the worst case -- near criticality -- scales quadratically with the number of qubits.
By pruning small parameters of the trained model, we find that the number of weights can be significantly reduced 
while still retaining an accurate reconstruction. This provides evidence that over-parametrization of the RBM is required to facilitate the learning process.
\end{abstract}

\maketitle

\section{\label{sec:level1}Introduction}

Generative models are a powerful class of machine learning algorithms that seek to reconstruct an unknown probability distribution $p(\mathbf{x})$ from a set of data $\mathbf{x}$.  
After training, generative models can be used to estimate the likelihood of new data not contained in the original set, or to produce new data samples for various purposes.
Recently, industry-standard generative models have been repurposed by the physics community with the goal of reconstructing a quantum wavefunction from projective measurement 
data \cite{torlai_learning_2016, torlai_Tomo, carrasquilla_povm}.
The question of scalability is of paramount importance for the reconstruction of quantum states prepared by near-term hardware which comprises tens or hundreds of qubits.

While several generative modeling techniques are available for quantum state reconstruction, by far the most well-studied  
involves restricted Boltzmann machines (RBMs) \cite{torlai_learning_2016,torlai_Tomo,RMBTN,BornM,DeepRBM}.
RBMs can be used to explicitly parametrize a probability distribution $p(\mathbf{x})$, and, through a suitable complex generalization, a quantum wavefunction \cite{torlai_Tomo,Carleo}.  
One main application of RBMs is the data-driven reconstruction of experimental states,
which has recently been demonstrated for a Rydberg-atom quantum simulator \cite{Rydberg}.
These and other uses have been covered extensively in the literature, including several recent reviews \cite{ML_NISQ,Perspective, MLPhysicsRev}.

With the steady increase in the size of experimental quantum devices,
an important question is how data-driven quantum state reconstruction scales with the number of qubits.
While many results have been reported for fixed finite-size reconstructions, less work has been done in the way of scaling analyses~\cite{neuromorphic}.  
Particularly important is the difference in scaling complexity of approximate machine learning methods for practical reconstructions,
as compared to full quantum state tomography that in general scales exponentially \cite{Tomobook}.

In this paper, we present a systematic study of the scaling of the computational resources required for 
accurate reconstruction of a quantum state. In particular, we focus on RBMs used to reconstruct 
the ground-state wavefunction of a one-dimensional transverse-field Ising model, which has a positive-real representation. 
Our training data is a set of projective measurements
sampled independently from a simulated tensor-network wavefunction.
We define a learning criterion based on the accuracy of the energy estimator of the RBM.
The state reconstruction is considered successful when the relative error of the energy estimator is smaller than a fixed threshold. We target in particular two contributions to the asymptotic scaling behavior in the many-qubit limit: the representational power of the neural network, i.e.,~the {\it expressiveness} of the parameterization of the state, and the amount of data required to train the model, also known as the {\it sample complexity}.

We find that deep within the ferromagnetic and paramagnetic phases, the number of RBM parameters required for accurate representation 
of the ground state is $\mathcal{O}(1)$.
As the transverse field is varied to approach the quantum critical point between these two phases, the
state becomes more challenging to reconstruct, as expected due to long-range quantum correlations that arise there.  
At the critical point, we observe that under standard RBM training procedures
the number of parameters grows quadratically in the number of qubits, $\mathcal{O}(N^2)$. 
The minimum number of measurements required to train this number of parameters scales linearly with the number of qubits, $\mathcal{O}(N)$.
Interestingly, we find that the number of parameters required for an accurate reconstruction can be significantly reduced post-training by pruning small weights and 
fine-tuning the RBM by a small number of additional training iterations.  We argue that an RBM 
requires over-parameterization to facilitate the optimization procedure associated with learning.

\section{\label{sec:level3} Defining a scaling study}

We are interested in probing the asymptotic scaling of the computational resources required to reconstruct a quantum state using an RBM.
The training set comprises projective measurement data produced from the ground-state wavefunction of the one-dimensional transverse-field Ising model (TFIM) defined by the Hamiltonian
\begin{equation}
H  = -J \sum_{\langle ij \rangle} \sigma^z_i \sigma^z_j - h \sum_i \sigma^x_i, \label{TFIMham}
\end{equation}
where ${\sigma^{x,y,z}}$ are Pauli operators, defined over $N$ sites (or qubits), and $\langle i j \rangle$ denotes nearest-neighbor pairs on a one-dimensional lattice
with open boundary conditions.
This model is thoroughly studied in the condensed matter and quantum information literature, and serves as a standard benchmark for many numerical methods, such as
quantum Monte Carlo \cite{SSETFIM,EstelleDMC}, Tensor Networks (TNs) \cite{MERA}, or more recent quantum optimization algorithms \cite{Hsieh1, Hsieh2, VITA}.  
We generate training data from a density matrix renormalization group (DMRG) simulation~\cite{Ferris} for various values of $h/J$ using the ITensor library \cite{ITensor}. The measurements of the ground-state wavefunction are produced in the $\sigma^z$ basis.

The Perron-Frobenius theorem guarantees that when the Hamiltonian Eq.~\eqref{TFIMham} has negative off-diagonal matrix elements in the $\sigma^z$ (computational) basis, the ground-state wavefunction is positive-real. Thus, there is a direct mapping between the wavefunction and a probability distribution, $\psi(\bm{\sigma}) = \sqrt{p(\bm{\sigma})}$.
This allows for a significant simplification in the RBM network structure, since complex phases or signs need not be parametrized.
In addition, the computational basis is trivially informationally complete, enabling training from data produced only in the $\sigma^z$ basis \cite{torlai_Tomo}.  

\subsection{Restricted Boltzmann machine}

The RBM consists of two layers of binary variables $v_i,h_j \in \{0,1\}$. The energy associated with each configuration is given by,
\begin{equation}
E_{\bm{\lambda}}(\bm{v},\bm{h})=-\sum_{ij}W_{ij}v_i h_j-\sum_i^N b_i v_i-\sum_j^{N_h} c_jh_j\: , \label{energy}
\end{equation}
where $N$ is the number of visible units, representing the qubits or spins, and $N_h$ is the number of hidden units parametrizing the interactions. The two layers are fully connected via the weight matrix $\bm{W}$ that, along with the bias terms $b_i$ and $c_j$, forms the set of learnable parameters $\bm{\lambda}=(\bm{W},\bm{b},\bm{c})$. The energy function Eq.~\eqref{energy} defines the joint probability distribution 
\begin{equation}
p_{\bm{\lambda}}(\bm{v},\bm{h})= \frac{1}{Z_{\bm{\lambda}}} e^{-E_{\bm{\lambda}}(\bm{v},\bm{h})},
\end{equation}
where $Z_{\bm{\lambda}}$ is the partition function of the machine.
The marginal distribution is obtained by tracing out the hidden units,
\begin{equation}
p_{\bm{\lambda}}(\bm{v})=\sum_{\bm{h}}p_{\bm{\lambda}}(\bm{v},\bm{h})=
\frac{1}{Z_{\bm{\lambda}}}\sum_{\bm{h}}e^{-E_{\bm{\lambda}}(\bm{v},\bm{h})} .
\end{equation}
It is this marginal distribution that forms the approximate representation of the ground state, $\psi_{\bm \lambda}(\bm{v}) = \sqrt{p_{\bm{\lambda}}(\bm{v})}$. 
In other words, because of the assumed positive-real form of the wavefunction, the training procedure is equivalent to conventional unsupervised learning of an RBM \cite{Hinton85}.
In particular, the objective of the training procedure is to minimize the Kullback-Leibler (KL) divergence, which defines the discrepancy between the distribution of projective measurements and the probability distribution parameterized by the RBM, through a method known as contrastive divergence \cite{HintonCD}.
In the present work, we use the QuCumber software package to implement and train a positive-real RBM \cite{qucumber}.

\subsection{Learning criterion}

In order to quantify the resources required for the data-driven reconstruction of the ground-state wavefunction for the TFIM, 
one must be able to assess when the learning is ``complete''.
Generally, the fidelity is considered a standard measure of the closeness of two quantum states, such as a target state and an approximate reconstructed state.  
However, in more generic situations than ours, where a TN representation of the target quantum state may not be available,
calculations of the fidelity typically scale exponentially, which renders them intractable for even moderate numbers of qubits.  
An alternative method for defining the accuracy of a reconstruction is to measure expectation values of local observables.  
Such expectation values can be efficiently calculated through standard estimators from samples produced by the RBM.
Importantly, these can be compared with the exact values measured from our DMRG simulations, or through other methods such as QMC that do not admit an explicit representation of the ground state.

The relative error between an RBM estimator and an exact DMRG expectation value will be referred to as the relative observable error (ROE).
For the current study, we define the {\it learning criterion} through the ROE in the expectation value of the energy, which can be calculated from the RBM using standard Markov Chain Monte Carlo techniques.
Take $\bar{U} = \langle H \rangle_{\rm RBM}$ to be the average of the energy estimator calculated from $n$ samples generated by the RBM.
Since $n$ is finite, a statistical error exists in the estimator, quantified by the standard deviation $\sigma$.
To account for this in a relative error measure, we compute the Gaussian confidence interval given by $\bar{U} \pm C \frac{\sigma}{\sqrt{n}}$.
The value of $C=2.576$ corresponding to $99\%$ confidence will be used throughout this paper. 
If $U = \langle H \rangle_{\rm exact}$ is the exact value of the energy estimator (calculated, e.g.,~with DMRG), then we can upper-bound the ROE by the larger relative error value of the confidence interval:
\begin{equation}
\epsilon = \max\left | \frac{U - (\bar{U} \pm C\sigma\sqrt{n}))}{U} \right |.
\end{equation}
Essentially, this means that we consider the learning to be ``complete'' when our desired upper bound on the ROE is satisfied 99\% of the time on our sample size.
We find empirically that $\epsilon = 0.002$ is a reasonable value that can be achieved by RBMs trained on TFIM data with
conventional algorithms for $N \le 100$ qubits. At smaller values (e.g.,~$\epsilon = 0.001$) training becomes impractical for $N > 50$, while for larger values we observed that the results reported below remain qualitatively the same; thus, we use $\epsilon = 0.002$ in the remainder of the paper.

With this learning criterion, we analyze the scaling behavior of the RBM by controlling two variables: 
the number of model parameters per qubit and the number of training measurements $M$, i.e.,~the sample complexity.  
However, we note that, as typical in machine learning studies, many other variables exist that are related to network architecture, learning rates, batch size, etc.~-- referred to as {\it hyperparameters}. Here these hyperparameters were made consistent for all values of $h/J$ and all system sizes $N$.

\section{\label{sec:level4}Results}

\begin{figure}[t]
\includegraphics[width=0.85\columnwidth]{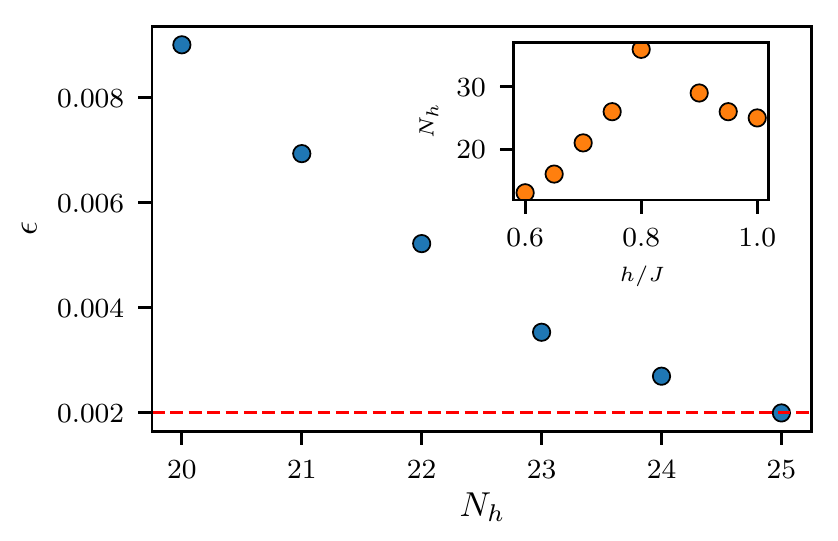}
\caption{The procedure used to determine the RBM expressiveness required to represent the
 TFIM wavefunction at $h/J=1$ with $N=50$ qubits. The number of hidden units $N_{h}$ is increased until the desired $\epsilon$ is achieved.
 The inset illustrates the number of hidden units required for convergence to $\epsilon \le 0.002$ for different values of $h/J$ near criticality. The position of the peak is discussed in the main text.
 }
\label{study1}
\end{figure}

In this section we present numerical results for the scaling of computational resources for reconstruction of the TFIM ground-state wavefunction for several values of $h/J$.  
In order to systematically investigate scaling, we control variables of interest in different ways as described in the following sections.

\subsection{Scaling of the model parameters} \label{Results1}

To begin, we are interested in the minimal number of RBM parameters per qubit required to faithfully reproduce the ground-state energy.
We parametrize this with the scaling of the size of the hidden layer, $N_h$.
We consider the critical point, corresponding to $h/J = 1$, as well as the ferromagnetic and paramagnetic phases.  
For each value of $N$, we produce large numbers of projective measurements
of $\sigma^z$ values using the DMRG simulation of the TFIM.  
Then, effectively assuming that the number of available training samples 
$M \rightarrow \infty$, we increase the number of hidden units $N_h$ until the learning criterion is uniquely satisfied for each value of $N$.

\begin{figure}[t]
\includegraphics[width=0.85\columnwidth]{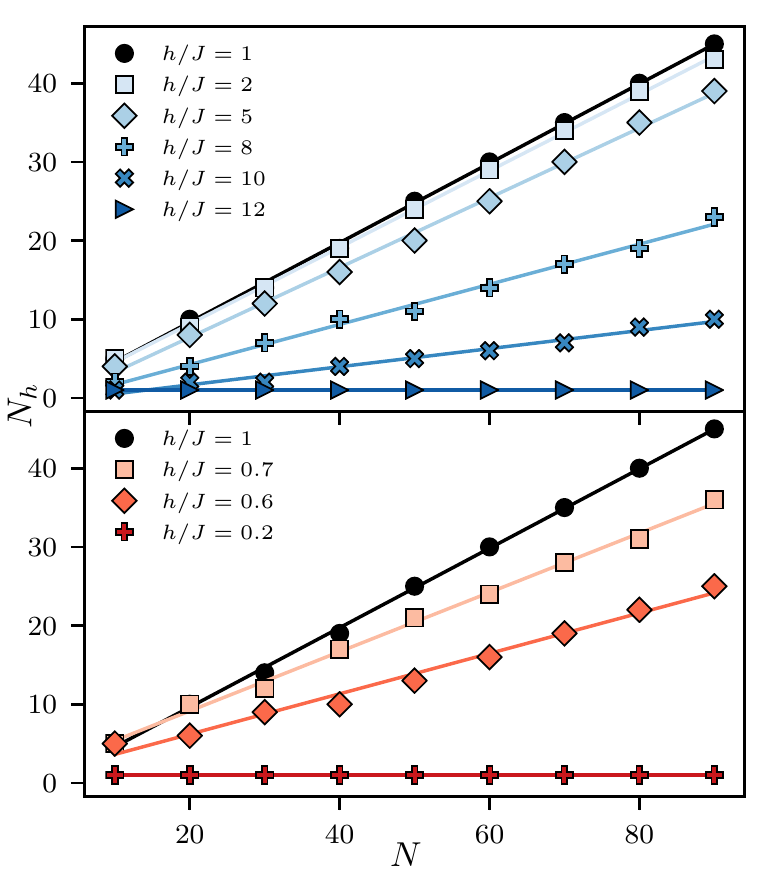}
\caption{Minimum number of hidden units $N_{h}$ required for $\epsilon \leq 0.002$ for various values of $h/J$. Straight lines are fits to the data.
}
\label{study2}
\end{figure}

Our procedure is illustrated in Fig.~\ref{study1} for a fixed system size of $N$ = 50.
In the main plot, corresponding to $h/J = 1$, we observe that the specified learning criterion $\epsilon = 0.002$ can not be achieved for $N_h < 25$.  
The minimum number of parameters required to accurately represent the ground-state wavefunction is thus $N_h=25$.
The inset illustrates the dependence of $N_h$ on field values near $h/J=1$, where the quantum wavefunction is most entangled.
One would expect that in the limit $N \rightarrow \infty$ the number of parameters required to accurately parameterize the 
wavefunction would be maximal at $h/J=1$. Curiously, we find that this peak occurs around $h/J \approx 0.8$, slightly on the
ferromagnetic side from the critical point. We hypothesize that this feature might be tied to the magnetization of the underlying dataset used for training, which was produced by our DMRG simulations in ITensor. For the maximum bond dimension that we employ (2000), the expected $\mathbb{Z}_{2}$ symmetry is not realized below certain values of the transverse field strength $h$ when the number of qubits is large. Furthermore, a similar phenomenon has been observed previously in studies of the relative energy in
diffusion Monte Carlo \cite{EstelleDMC} and a recent variational imaginary time ansatz \cite{VITA}. It would be an interesting topic of future study.

The result of repeating the above procedure for various numbers of qubits $N$ is illustrated in Fig.~\ref{study2}.  
For values of $h/J$ deep within the ferromagnetic or the paramagnetic phase, 
the required minimum number of hidden units scales as $N_h \sim \mathcal{O}(1)$ in the asymptotic limit of large $N$.  
This reflects the informational simplicity of the dataset close to the ferromagnetic or paramagnetic limits.
Near $h/J \approx 1$ the scaling of $N_h$ is clearly linear, meaning that the leading asymptotic scaling of the number of parameters
is $\mathcal{O}(N^2)$, as each additional hidden unit quadratically scales the number of parameters in the weight matrix $\bm{W}$.
In fact, for linear scaling in $N_h$ we expect a sub-leading term that scales proportional to $N$, due to the presence of the bias terms in Eq.~\ref{energy}.
However as noted in the Appendix, for data sets with an underlying $\mathbb{Z}_{2}$ symmetry, these bias terms do not represent independent parameters 
for the purpose of wavefunction reconstruction.
Finally, we note that for larger ROE thresholds $\epsilon > 0.002$ the prefactors and slopes are different, but 
the asymptotic scaling of the number of hidden units still remains linear near criticality.

\begin{figure}[t]
\includegraphics[width=0.85\columnwidth]{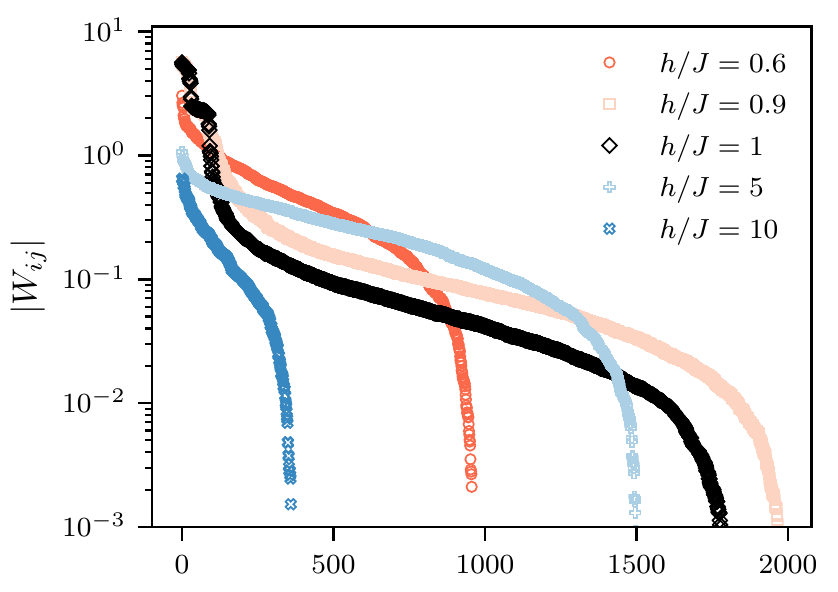}
\caption{Weight magnitudes, sorted in descending order from left to right, for various transverse field values and $N = 60$. Converged RBM models from the parameter study shown in Fig.~\ref{study2} are used here.}
\label{Wdistrib}
\end{figure}

Further insight on the above result can be obtained by looking at the distribution of the absolute weight values $| W_{ij} |$ in a typical trained model. In Fig.~\ref{Wdistrib} we plot the
magnitude of each individual weight, sorted in decreasing order from left to right, on a logarithmic scale.  
One can see that near criticality, the largest contribution is given by the first $10-20\%$ of weights; then 
the weight values decrease exponentially in magnitude, eventually falling off even more rapidly. We return to this observation after the next section.

\subsection{Scaling of sample complexity}

\begin{figure}[t]
\includegraphics[width=0.85\columnwidth]{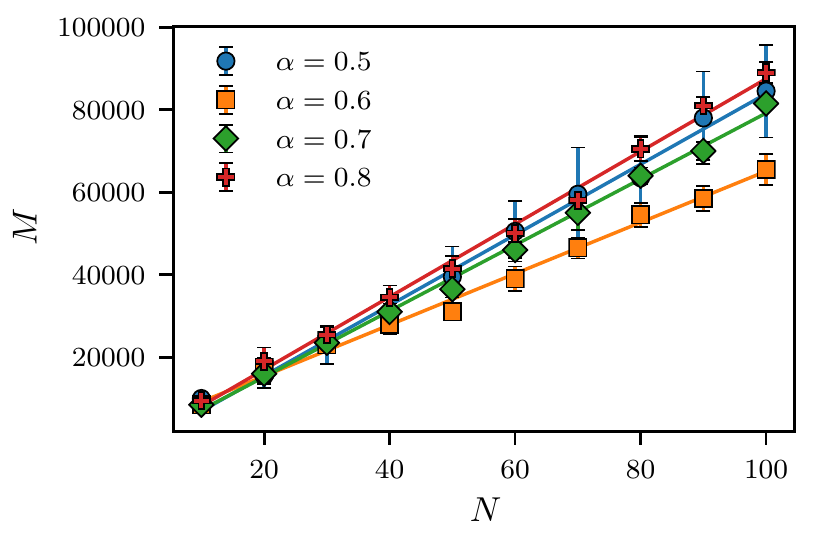}
\caption{The minimum number of training examples $M$ required for $\epsilon \leq 0.002$ for the TFIM at the critical point $h/J=1$ 
for different ratios of the number of hidden to visible units $\alpha = N_h/N$.
}
\label{studyM}
\end{figure}

Above, we studied the minimum number of RBM parameters required to find an accurate ground-state energy, assuming access to an infinite amount of training data. We now determine the minimal sample complexity required to accurately train this number of parameters. We focus on $h/J = 1$, and fix the ratio $\alpha = N_{h}/N$ for several values near $1/2$.  
Then, repeating the procedure from the last section, we increment the number of training examples $M$ by 2500 until the ROE learning criterion $\epsilon \le 0.002$ on the RBM energy estimator is achieved. This procedure is repeated for a number of different initial weight configurations, and the results are averaged. The resulting
scaling of the sample complexity is shown in Fig.~\ref{studyM}.

The results suggest that for $N_h/N$ near $1/2$, the sample complexity scales linearly in the number of qubits.  
Combining the asymptotic scaling results from the previous two sections, $N_h \sim N$ and $M \sim N$, 
suggest that the number of samples per parameter required to train a minimally-expressive machine scales as $N/N^2 \sim 1/N$.  
In other words, the relative ``data cost'' required to train a new weight parameter decreases with an increasing number of qubits.
A linear scaling of the sample complexity was also observed in a recent generative modeling scheme based on positive operator-valued measurements \cite{carrasquilla_povm}.

We remark that a sample complexity linear in $N$ is consistent with observations on the PAC-learnability of quantum states.  In Ref.~\onlinecite{learnQS} 
Aaronson argues that, 
if one is only concerned about learning a state well enough to predict the outcomes of most measurements drawn from it, 
the exponential cost usually associated with full state tomography is reduced to a linear scaling in $N$.  
This is what we find in Fig.~\ref{studyM}.
Indeed, a characteristic of the Aaronson  
learning theorem is the assumption that the training samples are drawn {\it independently} from the probability distribution. This is exactly the setting
that we employ in training the RBM in the present work. Hence it is reasonable to expect the theorem to apply.

\subsection{Reducing the number of model parameters post-training}

We now return to the results of Section~\ref{Results1}, where it was found that the minimal number of hidden units required to satisfy our chosen ROE scales
approximately as $N_h \approx \frac{1}{2} N$ near the TFIM quantum critical point.  
Implicit in this result is the RBM optimization procedure used to train the machine: a stochastic gradient descent 
that minimizes the KL-divergence \cite{ML_NISQ}. Since it is not obvious that the scaling behavior is
independent of this optimization procedure, it is fair to ask the question: Is it possible to find more efficient representations -- with fewer model parameters -- by modifying the learning protocol?
Indeed, it is known that the required number of model parameters is intertwined with  the specifics of the training procedure.
In particular, it has been found that the over-parameterization inherent to deep neural networks can ease and accelerate their
optimization by (stochastic) gradient descent \cite{nonconvOverparam, Livni14, Arora18, OverparamConvergence, OverparamSGD}.

Figure \ref{Wdistrib} offers a clue that the RBM parameterization may not be optimal (i.e.,~minimal) for the final trained wavefunction by demonstrating that the distribution of the weight magnitudes in a trained model is very non-uniform: 10 to 20\% of the weights have values that are orders of magnitude larger than the rest. Recent machine learning literature has studied the relative importance of these smaller weights with a procedure called {\it pruning}.
Following the ideas of Refs.~\cite{pruning,pruning2}, we define a pruning procedure for our scaling study in the following steps:
\begin{enumerate}
\item Start from the original, converged trained model (e.g., Fig.~\ref{Wdistrib}, with $N_h =  \frac{1}{2} N$ for $h/J = 1$).
\item Set a threshold $\delta$ for the weight magnitudes. If a given $|W_{ij}| < \delta$, set $W_{ij} = 0$, and freeze it for the following steps.
\item Fine-tune the pruned model by running several more training iterations until the desired accuracy (as defined by the ROE learning criterion) is restored.
\item Repeat steps 2-4, pruning additional weights until the model fails to fulfill the learning criterion.
\end{enumerate}
We choose the pruning threshold such that 40\% of the non-zero weights are pruned in the first iteration, and 5\% of the non-zero weights in each following iteration.
Note also that in this procedure we do not prune biases (see the Appendix for further comments).

\begin{table}[t]
\begin{tabular}{c|cccc}
\hline \hline
\( N \)& 10 & 20 & 30 & 40   \\
 \hline
original & 50 & 200 & 420 & 760   \\
pruned & 20 & 50 & 79 & 119 \\
\end{tabular}
\caption{The number of weights required to achieve $\epsilon \leq 0.002$ at the critical point $h/J = 1$. Results for the ``original'' RBM are taken from Fig.~\ref{study2}.}
\label{T1}
\end{table}

We apply weight pruning to our trained RBM focusing on the critical point of the TFIM, and
find that a significant reduction in the number of RBM parameters required to correctly capture the critical TFIM ground-state energy can be achieved for all system sizes. The results for several small numbers of qubits are presented in Table~\ref{T1}.  
We interpret this to mean that the standard training of an RBM with contrastive divergence benefits from an over-parameterization,
employing more weights than is strictly required for accurate expression of the TFIM wavefunction in order to make the optimization more navigable.
We note that in some rare cases, pruning a very small number of weights seriously alters the ROE,
highlighting that some paths through the optimization landscape may depend on weight parameters
that are not redundant.
For this reason, rigorous uncertainty intervals on our results are difficult to estimate at present.

The success of the pruning procedure opens up the possibility of systematically searching for a change in scaling behavior.
However, due to the significant increase in methodological complexity introduced by the pruning procedure, 
this analysis is out of scope for the current study and will be presented in another work.

\section{\label{sec:level5}Discussion}

In this paper, we have empirically studied the scaling of computational resources required for the accurate
reconstruction of positive-real wavefunctions using generative modeling with a restricted Boltzmann machine (RBM).
We obtained scaling results by examining the energy estimator calculated from an RBM
after training on projective measurement data from the one-dimensional transverse-field Ising model (TFIM) ground state.
An RBM reconstruction of the ground-state wavefunction was defined to be ``accurate'' when the relative error between the 
RBM estimator and the exact energy value was below a fixed threshold. Thus, scaling results in this paper are subject to the caveat that they 
could change if other criteria were to be considered, such as the convergence of fidelity or correlation functions.  

In the present case, convergence of the relative error in the energy produces several interesting results.  
First, for a standard optimization procedure with contrastive divergence, the number of weight parameters required for accurate 
reconstruction is at best constant (deep in the ferromagnetic/paramagnetic phases).  
At worst, this scaling is quadratic; this occurs near the quantum critical point between the two phases.  
In addition, the minimum number of samples required to converge the energy at the critical point is observed to scale linearly with 
the number of qubits.  This is consistent with a theorem by Aaronson that predicts a linear scaling in a similar setting for 
PAC-learning \cite{learnQS}.

Further, we present evidence that the number of parameters required to represent the ground state is drastically affected by
the RBM learning procedure.  By employing a pruning technique that sets small weights to zero, then fine-tuning the remaining
model parameters through additional training, we observe a very significant reduction in the number of parameters required to 
accurately reproduce the energy. 
It would be interesting to examine whether the asymptotic scaling functions identified here are affected by the pruning procedure.
Further, such a technique could provide a systematic way of searching for the minimal model expressiveness required for a given quantum state.
It would then be interesting to compare the obtained results to theoretical expectations for the representational capacity of RBMs required for quantum ground-state wavefunctions \cite{RMBTN,Gao17,NNQS}.

Indeed, numerous recent results have highlighted the benefit that over-parameterization provides for optimizing deep learning models \cite{OverparamConvergence, OverparamSGD, nonconvOverparam}. In this paper, we have discovered that 
RBMs trained on measurement data for positive-real wavefunctions may as well be aided by over-parameterization – beyond what is 
needed for the theoretical representation of the quantum state – as a means of assisting the standard optimization procedure 
of minimizing the KL divergence via contrastive divergence. The question of how to systematically mitigate this over-parameterization 
while still maintaining the ease of optimization is an active area of research \cite{LotteryTicket, DeconstructingLotto, SNIP}, one whose successes will be of great use for more efficiently representing and studying quantum systems.

It is natural to wonder what the scaling of computational resources is for reconstructing quantum states that are not real or positive. 
This question is especially pertinent for state-of-the-art experiments, such as fermionic quantum simulators~\cite{Mazurenko17},
wavefunctions generated by quantum dynamics~\cite{Lanyon2017,Keesling2018}, or quantum chemistry calculations with superconducting circuits~\cite{ibm_vqe}.
In contrast to positive wavefunctions, the reconstruction (with a suitably modified RBM) demands training data from an extended set of measurement bases. The ability to 
theoretically identify the minimal set, and how the size of this set scales with the number of qubits, will ultimately determine the feasibility of integrating this 
type of machine learning technology into such near-term quantum devices.  

In conclusion, we have proposed a systematic procedure to evaluate the scaling of resources for reconstructing positive-real wavefunctions with RBMs. 
A tighter threshold in the reconstructed energy accuracy, 
or improved neural-network parametrizations of non-positive states, will likely require a more powerful breed of generative model. Recurrent neural networks, 
transformers, and other autoregressive models are currently being considered in this context. 
In light of the fact that current intermediate-scale quantum devices are already capable of producing training data on 
tens and even hundreds of qubits, we expect these and similar scaling studies will be pursued in earnest in the near future.

\subsection*{Acknowledgements }
We acknowledge enlightening discussions with 
M. Beach,
J. Carrasquilla,
I. De Vlugt,
M. Ganahl,
E. Inack,
D. Kong,
E. Merali,
A. Rochetto
and D. Sels.
The DMRG calculations were performed using the ITensor libraries \cite{ITensor}.
The RBM calculations were performed with the QuCumber package \cite{qucumber}. 
This work was made possible by the facilities of the Shared Hierarchical Academic Research Computing Network (SHARCNET) and Compute Canada.
AG is supported by NSERC. RGM is supported by NSERC, the Canada Research Chair program, and the Perimeter Institute for Theoretical Physics. Research at Perimeter Institute is supported in part by the Government of Canada through the Department of Innovation, Science and Economic Development Canada and by the Province of Ontario through the Ministry of Economic Development, 
Job Creation and Trade. The Flatiron Institute is supported by the Simons Foundation.

\bibliographystyle{apsrev4-1}
\bibliography{bibliography.bib}

\subsection{Appendix}

Projective measurement data in the $\sigma^z$ basis for the TFIM will generally obey $\mathbb{Z}_{2}$ symmetry in the absence of symmetry breaking,
which may occur due to a limited DMRG bond dimension (or can happen spontaneously in the thermodynamic limit).
Let us assign the measurement on a single qubit $\sigma^z_i = \sigma_i = \pm 1$.
The probability of any given state over $N$ qubits, $\bm{\sigma}$, is therefore the same as that of the corresponding spin-flipped state $\bar{\bm{\sigma}}$, i.e.,~the magnetization of the state will be zero.
In a typical RBM, the values of the visible and hidden units will be 0 and 1, which can be mapped to an ``occupation'' number rather than a spin magnetization.
One can always consider instead the spin language by working in the $\pm{1}$ basis.
In this basis, when the underlying data set has zero magnetization, we can 
assume that the energy of the RBM takes the form 
\begin{equation}
E_{\bm{\lambda}}(\bm{\sigma}^v,\bm{\sigma}^h) = -\sum_{ij}{\Tilde{W}_{ij}\sigma^v_{i}\sigma^h_{j}}, \label{Senergy}
\end{equation}
where $\sigma^v_i$/$\sigma^h_j$ is a single visible/hidden unit in the spin language.
It can then be shown that this energy function, when used to define a joint distribution, results in 
$p_{\bm{\lambda}}(\bm{\sigma}^v) = p_{\bm{\lambda}}(\bar{\bm{\sigma}}^v)$ after marginalizing over the hidden units. 
In other words, the RBM in the $\pm1$ representation requires no biases (or magnetic fields) to capture a $\mathbb{Z}_{2}$ invariance in the data distribution. 

To see what this means in the occupation number representation, one can map $\sigma^v_{i} = 2v_{i} - 1$ and $\sigma^h_{j} = 2h_{j} - 1$ where ${v}_{i}$ and ${h}_{j} \in \left \{ 0,1 \right \}$. By transforming the energy expression Eq.~\eqref{Senergy}, and setting
${W}_{ij} = 4\Tilde{W}_{ij}$, ${b}_{i} = -\sum_{j}2\Tilde{W}_{ij}$ and  ${c}_{j} = -\sum_{i}2\Tilde{W}_{ij}$, we obtain an expression
identical to Eq.~\eqref{energy}.  This allows us to interpret the presence of biases, which are learned by the RBM even 
in TFIM data sets that are observed to be $\mathbb{Z}_{2}$ invariant.

\begin{figure}[b]
    \includegraphics[width=\columnwidth]{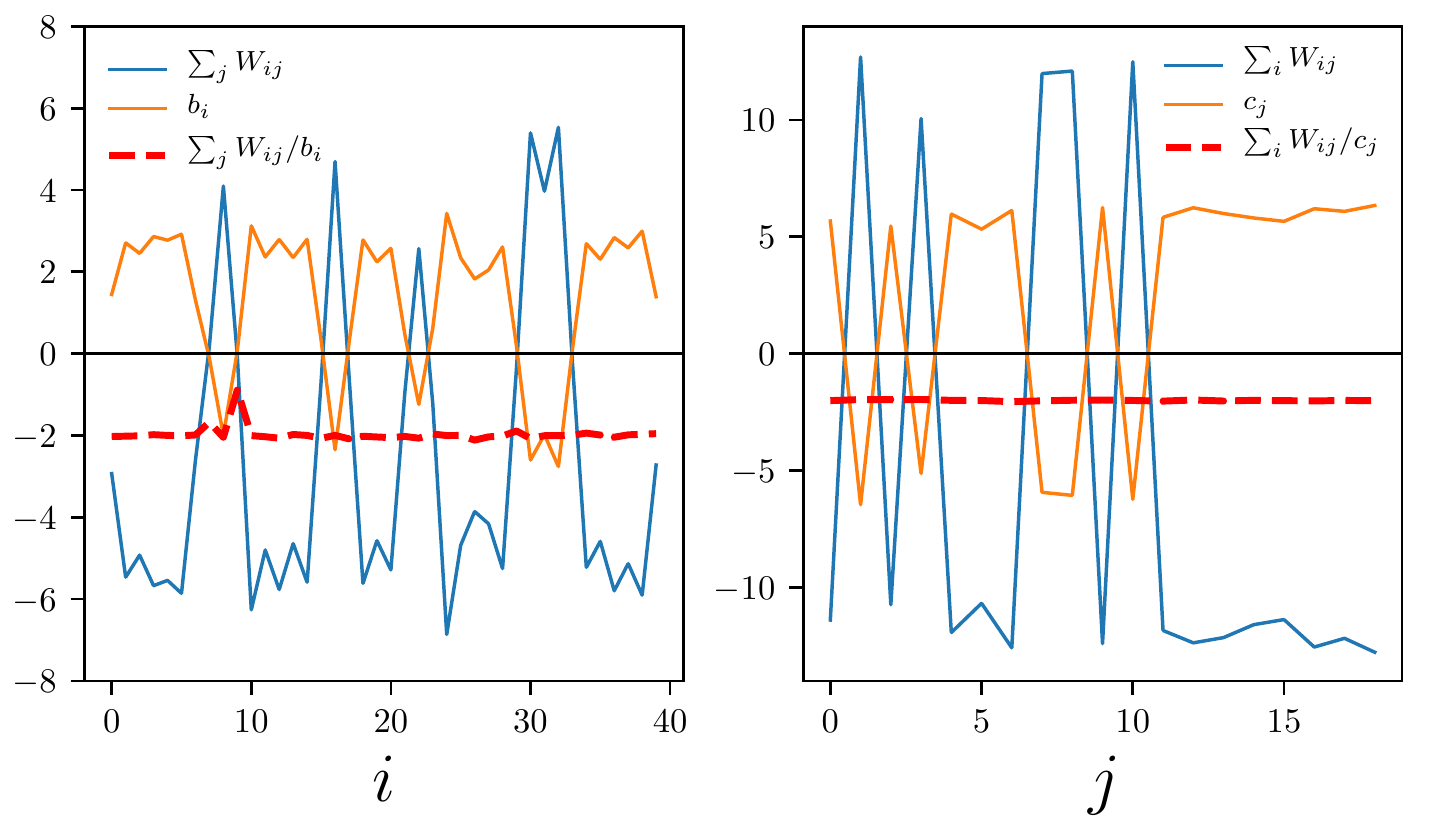}
    \caption{The left and right figures show the values of the weight matrix from an RBM trained on $N = 40$ at the quantum critical point to $\epsilon \leq 0.002$, summed along its rows and columns respectively. One can see that the ratios of the summed weights to biases fluctuate around -2, as predicted.}
    \label{fig:appfig}
\end{figure}

Let us examine the weight matrix of a typical converged run of the RBM.
From the above arguments, can calculate the ratios of the biases to the sums of weights along rows and columns of the weight matrix:
\begin{equation}
    \alpha_{i} = \sum_{j}{{W}_{ij}}/{b}_{i} = -\sum_{j}{4\Tilde{W}_{ij}}/\sum_{j}{2\Tilde{W}_{ij}} = -2,
\end{equation}{}
\begin{equation}
    \beta_{j} = \sum_{i}{{W}_{ij}}/{c}_{j} = -\sum_{i}{4\Tilde{W}_{ij}}/\sum_{i}{2\Tilde{W}_{ij}} = -2.
\end{equation}{}
Thus one expects at least approximately that $\alpha_i = \beta_j = -2$ for all $i$ and $j$ for any of the trained RBM models considered in Fig.~\ref{study2}.   We confirm this 
behavior for $N=40$ and $N_h = 20$ in Fig.~\ref{fig:appfig}.

{}

\end{document}